

\input phyzzx

\catcode`@=11
\newtoks\KUNS
\newtoks\HETH
\newtoks\monthyear
\Pubnum={KUNS~\the\KUNS\cr HE(TH)~\the\HETH}
\monthyear={September, 1992}
\def\p@bblock{\begingroup \tabskip=\hsize minus \hsize
   \baselineskip=1.5\ht\strutbox \topspace-2\baselineskip
   \halign to\hsize{\strut ##\hfil\tabskip=0pt\crcr
   \the\Pubnum\cr hep-ph/9210228\cr \the\monthyear\cr }\endgroup}
\def\bftitlestyle#1{\par\begingroup \titleparagraphs
     \iftwelv@\fourteenpoint\else\twelvepoint\fi
   \noindent {\bf #1}\par\endgroup }
\def\title#1{\vskip\frontpageskip \bftitlestyle{#1} \vskip\headskip }
%
%
\def\Kyoto{\address{Department of Physics,~Kyoto University \break
                            Kyoto~606,~JAPAN}}

%
%
\paperfootline={\hss\iffrontpage\else\ifp@genum%
                \tenrm --\thinspace\folio\thinspace --\hss\fi\fi}
\footline=\paperfootline
%
%

%
\def\journal#1&#2(#3){\begingroup \let\journal=\dummyj@urnal
    \unskip, \sl #1\unskip~\bf\ignorespaces #2\rm
    (\afterassignment\j@ur \count255=#3) \endgroup\ignorespaces }
\def\andjournal#1&#2(#3){\begingroup \let\journal=\dummyj@urnal
    \sl #1\unskip~\bf\ignorespaces #2\rm
    (\afterassignment\j@ur \count255=#3) \endgroup\ignorespaces }
\def\andvol&#1(#2){\begingroup \let\journal=\dummyj@urnal
    \bf\ignorespaces #1\rm
    (\afterassignment\j@ur \count255=#2) \endgroup\ignorespaces }
\def\NP{Nucl.~Phys. }

\def\PRL{Phys.~Rev.~Lett. }
\def\PL{Phys.~Lett. }
\def\PTP{Prog.~Theor.~Phys. }

%
%
%
\def\ee{\eqno\eq }
\KUNS={1161}     
\HETH={92/11}   

\def\1#1{{1 \over {#1}}}
\def\2#1{{{#1} \over 2}}

\def\calD{{\cal D}}

\def\smg{{\raise2.2pt\hbox{$>$}}\kern-8pt\lower3pt\hbox{$\sim$}}

\def\sml{{\raise2.2pt\hbox{$<$}}\kern-7pt\lower3pt\hbox{$\sim$}}
\def\siml{\ \sml \ }

\def\lbar{\overline}
\def\MS{${\lbar {\rm MS}}$\ }
\def\ep{effective potential}

\def\sm{{1 \over 2}\lambda\phi^2+m^2}

\def\der#1{{\partial \over \partial #1}}
\def\SUM#1#2{\sum_{\scriptstyle #1
                   \atop \lower2pt\hbox{$\scriptstyle #2$}}}

\catcode`@=12

\REF\GQW{
   H.~Georgi, H.~Quinn and S.~Weinberg
      \journal \PRL &33 (74) 451.
}
%

\REF\GRZ{
   G.~Gamberini, G.~Ridolfi and F.~Zwirner
      \journal \NP &B331 (90) 331.
}
\REF\BKMN{
   M.~Bando, T.~Kugo, N.~Maekawa and H.~Nakano,
\nextline
           Preprint KUNS 1129 (1992).
}
%
\REF\CW{
   S.~Coleman and S.~Weinberg
      \journal Phys. Rev. &D7 (73) 1888.
}
%
\REF\Kast{
   B.~Kastening
      \journal \PL &B283 (92) 287.
}
\REF\BKMNa{
   M.~Bando, T.~Kugo, N.~Maekawa and H.~Nakano,
\nextline
           Preprint KUNS 1162 (1992).
}
\REF\Kugo{
   T.~Kugo
      \journal Soryusiron Kenkyu &53 (76) 1,
\nextline
           see also  \andjournal \PTP &57 (77) 593.
}
\REF\Sher{
   M.~Sher
      \journal Phys. Rep. &179 (89) 273.
}
\REF\EJ{
   M.~B.~Einhorn and D.~R.~T.~Jones
      \journal \NP &B230[FS10] (90) 261.
}


\titlepage

\title{Improving the Effective Potential}

\author{   Masako~BANDO      }

\address{   Aichi University, Miyoshi, Aichi 470-02      }

\andauthor{   Taichiro~KUGO,\ \ \ Nobuhiro~MAEKAWA        }
\vglue0.1cm
\centerline{ and }
\vglue0.1mm
\titlestyle{\twelvecp    Hiroaki~NAKANO      }

\Kyoto

\abstract{
A general procedure is presented how to improve the
effective potential by using the renormalization group equation (RGE)
in
\MS scheme. If one knows the $L$-loop effective potential
and the RGE coefficient functions up to $(L+1)$-loop level,
this procedure gives
an improved potential which satisfies the RGE and
contains all of the leading, next-to-leading,
$\cdots $, and $L$-th-to-leading log terms.
 }

\endpage          

\sequentialequations 

Since the work of Georgi, Quinn and Weinberg,\refmark{\GQW}
 it is a standard procedure to
use renormalization group equation (RGE) to discuss the low-energy
physics,
in a system which is supposedly described by a certain unified theory
at a very large energy scale. In recent active investigations
in the minimal
 supersymmetric standard model, people also discuss the effective
potential
for the Higgs fields in which the coupling constants and masses
are running parameters which depend on the renormalization point
$\mu $.
It is quite legitimate to use such renormalized parameters
with renormalization point $\mu $ chosen to be a value of the order of
the energy
scale at which we discuss the physics.

It was, however, found that the tree
effective potential with such running parameters inserted is too
sensitive to
the choice of the renormalization point $\mu $: for instance, the
vacuum
expectation value (VEV) of the Higgs field rapidly varies
against a small change of the renormalization point,
so that no reliable
prediction can be made. To save this situation it was proposed to use
the \ep\ at the 1-loop level instead.\refmark{\GRZ}
 This indeed improved
the situation and the $\mu $-dependence, \eg, of the Higgs VEV, became
much
milder than in the tree case.
But, in some cases,\refmark{\BKMN} the stability against $\mu $
achieved
by this is not enough and the Higgs VEV
still shows a rapid $\mu $-dependence for
$\mu $ one-order apart from the supersymmetry breaking scale.

This is of course a problem which arises from the fact that the
used \ep\ itself is not satisfying the RGE:
the RGE differential operator is just a
total derivative $d/d\ln\mu $, so if the \ep\ satisfies the RGE,
it {\it is} $\mu $-independent
and the VEV realized as its minimum should show only
a very mild (logarithmic) $\mu $-dependence of the wave function
renormalization.
The purpose of this letter is to present a simple procedure to improve
\ep\
so as to satisfy the RGE.

It is a bit surprising why such a procedure has not been known for a
general
system. Coleman and Weinberg\refmark{\CW}
 are probably the first to improve the \ep\ using
RGE. Their method, however, relied on a special definition of the
$\lambda \phi ^4$
coupling constant as a fourth derivative of the potential, and was
restricted
to massless systems.  Very recently Kastening wrote a remarkable paper,
\refmark{\Kast}
in which he presented two method to obtain RGE improved \ep\
in massive
$\lambda \phi ^4$ theory: one method uses a detailed form of the \ep\
expressed as a
power series in several variables and determines the coefficients
of the series by inserting that form in the RGE.
This is probably too complicated to
extend it to more realistic systems. Another method is a smarter one
which
may have a possibility of generalization. However this method as it
stands
also has problems: it still expands the \ep\ in a power series in a
certain
variable and solves the RGE order by order, which is
in fact a complicate
procedure for a realistic system. Another problem is that he had to
make a
peculiar ansatz for the form of the vacuum energy (\ie,
$\phi $-independent) term
of the \ep. [It is  peculiar since it diverges when $\lambda $ goes to
zero.]

Our present work is in a sense a re-organization of his second method.
The above mentioned second problem has a simple solution: if we
properly take
the renormalization of the vacuum energy term into account, we do not
need any
special ansatz for it. [In any case it is very interesting that such
vacuum
energy term becomes relevant to the $\phi $-dependent terms of
the \ep.] As for the
first problem  we do not solve  the RGE for \ep\ order by order but
uses the
well-known full order solution itself. This is the main point of our
method,
which greatly simplifies the procedure and makes it possible to apply
to more general systems. Indeed we shall show
that this procedure applies
to any system which has essentially a unique mass scale. [Actually,
with a suitable modification, it is
extendible also to completely general system possessing many mass
scales, as
will be shown in a separate paper.\refmark{\BKMNa}]

To explain the essence of our procedure,
we consider the simplest case of $\lambda \phi ^4$ model of a real
scalar field.
The Lagrangian of this system is given by
$$
{\cal L}=
{1\over 2}(\partial \phi )^2 -{1\over 2}m^2 \phi ^2 -{1\over 4!}
\lambda \phi ^4 - hm^4 \ .
\eqn\eqmodel
$$
The last term $hm^4$ is the vacuum energy term which is
usually omitted.
But surprisingly, in the mass-independent renormalization scheme,
it becomes relevant to us in the calculation of \ep\ as
we shall see below. Renormalization of the vacuum energy term
is performed in the \MS scheme simply by omitting the `divergent' term
proportional to positive powers of
$1/\bar\varepsilon\equiv 2/(4-n)-\gamma +\ln4\pi $
in the calculated vacuum energy $V(\phi =0)$.  The corresponding
counter-terms
are supplied by the renormalization of
the bare vacuum-energy parameter $h_0 = Z_h \mu ^{4-n}h$, from
which the renormalized vacuum-energy parameter $h$ becomes dependent
on the
renormalization point $\mu $.
\foot{In the orthodox mass-independent renormalization \refmark{\Kugo}
with cutoff
regularization, one should impose the following
three renormalization conditions to renormalize the vacuum energy term
(0-point function) $\Gamma ^{(0)}$ as a function of $m^2$:
$$
\Gamma ^{(0)}(m^2)\Big\vert_{m^2=0}
=\der{m^2}\Gamma ^{(0)}(m^2)\Big\vert_{m^2=0}=0\ ,
\quad \Big(\der{m^2}\Big)^2\Gamma ^{(0)}(m^2)
\Big\vert_{m^2=\mu ^2}=-h\ .
$$
These are realized by counter-terms of the form $A+Bm^2+Cm^4$.
}

The \ep\ at the 1-loop level is calculated in \MS scheme as
$$
\eqalign{
V_1 &=V^{(0)}+V^{(1)} \ , \cr
V^{(0)} &= {1\over 2}m^2 \phi ^2 +{1\over 4!}\lambda \phi ^4 + hm^4
\ ,\cr
V^{(1)} &= {1\over 4\cdot 16\pi ^2}
 M_\phi ^4\Big(\ln{M_\phi ^2\over \mu ^2}-{3\over 2}\Big) \ ,\cr
}\eqn\eqep
$$
where
$$
M_\phi ^2\equiv {1\over 2}\lambda \phi ^2+m^2
\eqn\eqmass
$$
is the scalar mass in the presence of scalar background $\phi $.

Renormalization theory tells us that \ep\ satisfies RGE:
$$
{\cal D} V(\phi ,m^2,\lambda ,h;\,\mu ) = 0 \ ,
\eqn\eqRGE
$$
with
$$
\calD = \mu \der\mu +\beta \der\lambda -\gamma _mm^2\der{m^2}-
\gamma _\phi \phi \der\phi +\beta _h\der{h} \ .
\eqn\eqRGED
$$
The well-known solution is given in the form:
$$
V(\phi ,m^2,\lambda ,h;\,\mu ^2) =
V\big(\bar\phi (t),\bar m^2(t),\bar\lambda (t),\bar h(t); \,e^{2t}\!
\mu ^2\big)\ ,
\eqn\eqSOLUTION
$$
where $\bar\lambda $, $\bar m^2$, $\bar\phi $ and $\bar h$ are
running parameters whose $t$-dependence is determined by
$$
\eqalign{
{d \bar\lambda (t)\over dt} &= \beta \big(\bar\lambda (t)\big) \ ,\cr
{d \bar X(t)\over dt} &= -\gamma _X\big(\bar\lambda (t)\big) \,
\bar X(t) \qquad
               {\rm for}\ \ X = m^2, \phi , \cr
{d \bar h(t)\over dt} &= \beta _h\big(\bar\lambda (t), \bar h(t)\big)
\ ,\cr
}\ee
$$
with the boundary condition that they reduce to the unbarred
parameters at
$t=0$. Note that the vacuum-energy parameter $h$  can affect only the
evolution of itself.

The general solution \eqSOLUTION\ gives full information of RGE:
it says that, as a result of the fact
that RGE is a first order differential equation, the \ep\
is determined once its function form is known at a certain value of
$t$.
Namely, RGE reduces the number of variables on which the \ep\ depends
by one.
That is all. So, to derive  useful information from RGE,
we need to know anyway the function of \ep\ at a certain value of $t$,
a `boundary' or `initial' function.

Let us first see the logarithm structure of the \ep. We note that
the $L$-loop ($L\geq 1$) level contribution to the \ep\ has the
following form:
$$
V^{(L)} = \lambda ^{L-1} M_\phi ^4 \times
\big[ \hbox{polynomial in }\ \ln{M_\phi ^2\over \mu ^2}\ {\rm and}\
{\lambda \phi ^2\over M_\phi ^2} \big]
\ . \eqn\eqLLOOP
$$
This can be understood most easily in the following way:
To compute the
\ep\
$V(\phi )$, we first rewrite the quantum Lagrangian in the form
$$
{\cal L}= {1\over \lambda } \bigg[
{1\over 2}(\partial (\sqrt\lambda \Phi ))^2 -{1\over 2}m^2
(\sqrt\lambda \Phi )^2
-{1\over 4!}(\sqrt\lambda \Phi )^4 - \lambda hm^4 \bigg] \ ,
\eqn\eqLAG
$$
and then make the field shift $\Phi \rightarrow \Phi +\phi $ and
regard
$\sqrt\lambda \Phi $ as our basic
quantum field.
In this form the parameters characterizing the theory are only
the scalar mass $M_\phi ^2={1\over 2}\lambda \phi ^2+m^2$,
the cubic coupling $\sqrt\lambda \phi $
 and $\lambda $ (aside from the vacuum-energy term). Moreover the
last parameter $\lambda $ is no longer the quartic coupling constant
but an overall factor
in front of the action just like Planck constant $\hbar $.
Then  the above form of Eq.\eqLLOOP\  is self-evident.
Now let us introduce the following variables:
$$
\eqalign{
s &\equiv  \lambda \ln{M_\phi ^2\over \mu ^2}\ , \cr
x &\equiv  {{\lambda \over 2}\phi ^2\over M_\phi ^2} \ ,\cr
z &\equiv   \lambda  h {m^4\over M_\phi ^4}\ .\cr
}\eqn\eqVAR
$$
In terms of these variables together with $\lambda $, the 1-loop
potential
\eqep, for example, can be expressed as
$$
V_1  =  {M_\phi ^4\over \lambda } \Big[
x(1-x) +{1\over 3!}x^2 + z  + {1\over 64\pi ^2} \big(s-{3\over 2}
\lambda \big) \Big] \ .
\ee$$
Since we know that the logarithms appear only up to $L$-th power at
the $L$
loop level, the $L$-loop contribution \eqLLOOP\ takes the form
\def\vl#1{v^{(L)}_{#1}(x)}
$$
V^{(L)} = {M_\phi ^4\over \lambda } \times \big[ \vl{0} s^L + \lambda
\vl{1}s^{L-1}
    + \lambda ^2\vl{2}s^{L-2} + \cdots + \lambda ^L\vl{L} \big]  \ ,
\ee
$$
so that the full \ep\ has the form:
$$
\eqalignno{
 V &=  {M_\phi ^4\over \lambda } \sum_{\ell=0}^\infty  \lambda ^{\ell}
      \big[f_\ell(s,x)+z\delta _{\ell,0}\big]     &\eqname\eqLLOG  \cr
   &\equiv  M_\phi ^4 \widetilde V(s, x, z, \lambda )   \ ,      &
  \eqname\eqVTILDE \cr
 f_\ell&(s,x)
   =\sum_{L=\ell}^\infty \vl{\ell}s^{L-\ell}  \ .   &\eqname\eqFELL \cr
}$$
This form of expansion \eqLLOG\ in powers of $\lambda $,
which was first derived by Kastening,\refmark{\Kast}
just gives a {\it leading-log series expansion}: namely, the functions
$f_0, f_1, \cdots $ correspond to the leading, next-to-leading,
$\cdots $ log terms, respectively. So the explicit $\lambda $ factors
which appear
when the expression is written in terms of variables
$s, x, z$ and $\lambda $ show the
order in this leading-log series expansion. We refer to the term
proportional
to $\lambda ^{\ell-1}$ in $V$ as $\ell$-th-to-leading log term.
Unlike Kastening, we do not write the RGE for
the functions $f_\ell(s,x)$ separately
since we do already know the solution \eqSOLUTION\
to the RGE for the full \ep. This greatly simplifies the procedure
for the
practical use.

In this form, the above solution \eqSOLUTION\ of RGE is rewritten into
$$
V=M_\phi ^4\widetilde V(s,x,z,\lambda )
   = \overline M_\phi ^4(t)\,
   \widetilde V\big(\bar s(t), \bar x(t), \bar z(t), \bar\lambda (t)
   \big) \ ,
\eqn\eqSOLTHREE
$$
where the barred quantities $\overline M_\phi ^2(t),\ \bar s(t),\
\bar x(t)$ and $\bar z(t)$ are the
variables $M_\phi ^2,\ s,\ x$ and $z$ at `time' $t$: \eg,
$$
\eqalign{
\overline M_\phi ^2(t)\equiv &{1\over 2}\bar \lambda (t)\bar \phi ^2(t)
+\bar m^2(t)\ ,  \cr
\bar s(t) \equiv  & \bar\lambda (t)\ln{\overline M_\phi ^2(t)\over
e^{2t}\mu ^2} \ .\cr
}\eqn\eqSBAR
$$
Since this expression \eqSOLTHREE\
says that it is $t$-independent, we can put any $t$ and we
should look for such $t$ at which we can calculate the function form
of \ep.

Now we come to the point. The form \eqFELL\ tells us
that at $s=0$ the $\ell$-th-to-leading log
function $f_\ell$ is given solely
in terms of $\ell$-loop level potential:
$f_\ell(s=0,x)=v^{(L=\ell)}_{\ell}(x)
=\widetilde V^{(\ell)}(s=0, x, \lambda )/\lambda ^{\ell-1}$.
So, if we calculate the \ep\ up to
$L$-loop level, $V_L=V^{(0)}+V^{(1)}+\cdots +V^{(L)}$,
then at $s=0$ it already gives the function {\it exact}
up to $L$-th-to-leading log order:
$$
V=M_\phi ^4\widetilde V(s=0,x,z,\lambda )
=M_\phi ^4\widetilde V_L(s=0,x,z,\lambda ) + O(\lambda ^L) \ .
\ee$$
That is, we can use the function $V_L\big\vert_{s=0}$
as a `boundary' function required in the RHS of
the solution \eqSOLUTION\
or \eqSOLTHREE\ of RGE. Therefore, with the $L$-loop potential $V_L$
at hand,
the \ep\ satisfying the RGE can be given by
$$
\eqalign{
M_\phi ^4\widetilde V(s,x,z,\lambda ) &=
\overline M_\phi ^4(t)\,
\widetilde V_L\big(\bar s(t)=0, \bar x(t), \bar z(t), \bar\lambda (t)
\big) \cr
&= \overline M_\phi ^4(t) \sum_{\ell=0}^L \bar\lambda ^{\ell-1}(t)
      \Big[v^{(\ell)}_\ell\big(\bar x(t)\big)
      +\bar z(t)\delta _{\ell,0}\Big]_{\bar s(t)=0} \ ,\cr
}\eqn\eqSOLFOUR
$$
or, equivalently,
$$
V(\phi ,m^2,\lambda ,h;\,\mu ^2) =
V_L\big(\bar\phi (t),\bar m^2(t),\bar\lambda (t),\bar h(t); \,e^{2t}
\!\mu ^2\big)
\Big\vert_{\bar s(t)=0}\ .
\eqn\eqSOLTWO
$$
The barred quantities in these expressions
should of course be evaluated at $t$ satisfying $\bar s(t)=0$.

The process of solving $\bar s(t)=0$ with respect to $t$,
if one wishes,
 may be bypassed as follows.
Running of this variable $\bar s\equiv \bar s(t)$ is determined by the
differential equation:
$$
\eqalign{
{d\bar s \over dt}&= \beta _s(\bar s, \bar m^2, \bar \lambda ,
\bar \phi )
   \equiv  \bar \beta _s  \ ,\cr
\beta _s&=\lambda \Big[{\beta \over \lambda ^2}s-2\Big]
   + \lambda ^2\Big[\big({\beta \over \lambda ^2}-2{\gamma \over
\lambda }\big)x-{\gamma _m\over \lambda }(1-x)\Big] \ .\cr
}\eqn\eqBETAS
$$
We can switch to use the variable $\bar s$ itself
in place of the `time'
variable $t$, and then we regard the running quantities
$\bar\lambda , \bar m^2, \bar\phi $ and $\bar h$ as functions of
$\bar s$
(and of initial parameters $\lambda , m^2, \phi $ and $h$, of course).
Their runnings with respect to $\bar s$ are of course determined
by equations
$$
{d\bar X\over d\bar s}={1\over \bar \beta _s}{d\bar X\over dt}=
{\bar\beta _X\over \bar \beta _s}\ .
\eqn\eqBETA
$$
Then the quantities $\bar\phi (t),\bar m^2(t),\bar\lambda (t)$ and
$\bar h(t)$ in the
RHS of the solution \eqSOLTWO\ are simply obtained by setting
their argument $\bar s$ equal to zero.

Although the solution \eqSOLTWO\ is `exact'
only up to $L$-th-to-leading log order,
it satisfies the RGE {\it exactly}
if the runnings of the barred quantities are
solved exactly (,which is independent of the choice of the `boundary'
function). If the runnings of the parameters
$\bar \lambda /\lambda , \bar \phi /\phi , \bar m^2/m^2$ and
$\bar h/h$  are solved correctly only
up to {\it $L$-th power in } $\lambda $ in the sense of leading
log series expansion, our solution \eqSOLTWO\
satisfies the RGE up to $L$-th-to-leading log order and is `exact'
in that
order. This can be understood from the second expression in \eqSOLFOUR;
 if
$\bar \lambda /\lambda , \bar\phi /\phi , \bar x/x$ and $\bar z/z$ are
determined
correctly up to the $L$-th power in $\lambda $, their errors are of
the order
$O(\lambda ^{L+1})$ and
can change our \ep\ in \eqSOLFOUR\ at most by an $O(\lambda ^L)$
quantity\rlap.\foot{
It may be of help to give here a `quick table' showing
which quantities are of which order in $\lambda $
(as the expansion parameter of the leading-log series expansion):
$$
\bar\lambda (t) \ \sim \ O(\lambda ^1) , \quad \bar\phi (t) \ \sim
\ O(\lambda ^{-\1{2}}) , \quad
\bar m^2(t) \ \sim \ O(1) , \quad \bar h(t) \ \sim \ O(\lambda ^{-1})
\ .
$$
This follows from the RG running equation
and the fact that $m^2, \lambda \phi ^2$ and $\lambda  h$ are
of $O(1)$ in the leading-log series expansion.
}

To achieve this `exactness' up to $L$-th-to-leading
log order, it is sufficient to know the {\it $(L+1)$-loop
RGE coefficient functions } $\beta , \gamma $ and so on.
This is because we need those coefficient functions $\beta /\lambda ,
\ \gamma ,\ \gamma _m$
and $\beta _h/h$ correct up to $O(\lambda ^{L+1})$ (but not
$O(\lambda ^L)$),
since $\bar \beta _s$ in the RG running eq.\eqBETA\
with respect to $\bar s$
 is  of $O(\lambda ^1)$.
Thus,
{\it with $L$-loop \ep\ and $(L+1)$-loop RGE coefficient functions,
we can obtain an RGE improved \ep\
which is exact up to $L$-th-to-leading log order.}

That is all of our procedure: the Eq.\eqSOLTWO\ gives the final
answer of
our improved \ep.

Let us now demonstrate these processes by explicit computations
to the leading log order (\ie, $L=0$).
The coefficient functions of RGE are calculated at the 1-loop level as
$$
\eqalign{
\beta  &= {1\over 16\pi ^2}3\lambda ^2\equiv \beta _{1}\lambda ^2
\ ,\cr
\gamma _m &= -{1\over 16\pi ^2}\lambda  \equiv  \gamma _{m1}\lambda
\ ,\cr
\gamma  &= {1\over 16\pi ^2}\times 0 \equiv  \gamma _1\lambda  \ ,\cr
\beta _h &= +2h\gamma _m +{1\over 16\pi ^2}{1\over 2} \equiv   2h
\gamma _m +\beta _{h1} \ .\cr
}\ee
$$
Noting that  $\bar\beta _s = (\beta _1\bar s-2)\bar\lambda +
O(\lambda ^2)$,
the RGE for $\bar \lambda $
$$
\bar\beta _s{d\bar\lambda  \over d\bar s} = \bar\beta
\ee$$
gives in the lowest order in $\lambda $:
$$
(\beta _1\bar s-2)\bar\lambda {d\bar\lambda  \over d\bar s} =
\beta _1 \bar\lambda ^2 \ .
\ee$$
This is integrated as
$$
\int _\lambda ^{\bar\lambda } {d\bar\lambda  \over \bar\lambda } =
\int _s^{\bar s} {d\bar s \over \bar s - {2 \over \beta _1} } \ ,
\ee$$
and gives
$$
\bar\lambda  = \lambda {1-{\beta _1\over 2}\bar s \over  1-
{\beta _1\over 2}s } \ .
\ee$$
Similarly, we have
$$
\eqalign{
(\beta _1\bar s-2)\bar\lambda {d\bar\phi  \over d\bar s} &= -\gamma _1
\bar\lambda  \bar\phi  \ , \cr
(\beta _1\bar s-2)\bar\lambda {d\bar m^2 \over d\bar s} &=
-\gamma _{m1} \bar\lambda \bar m^2 \ ,\cr
}\ee
$$
so that we get
$$
\eqalign{
\bar\phi  &=
   \phi \left({1-{\beta _1\over 2}\bar s \over  1-{\beta _1\over 2}s }
\right)^{-{\gamma _1\over \beta _1}} \ ,\cr
\bar m^2 &=
   m^2\left({1-{\beta _1\over 2}\bar s \over  1-{\beta _1\over 2}s }
   \right)^{-{\gamma _{m1}\over \beta _1}} \ .\cr
}\ee
$$
Finally, instead of $\bar h$, we write the RGE
for $\bar h\times \bar m^4$ which reads
$$
\eqalign{
(\beta _1\bar s-2)\bar\lambda {d(\bar h\bar m^4) \over d\bar s} &=
 \bar m^4
\big[\bar \beta _h -2\bar \gamma _m h \big]_{\hbox{lowest order in }
\lambda }  \cr
 &= \beta _{h1}\bar m^4 \ .  \cr
}\ee
$$
Note that the $2h\gamma _m$ part in $\beta _h$ is cancelled.
We obtain
$$
\eqalign{
\int _{hm^4}^{\bar h \bar m^4}d(\bar h\bar m^4)
 &= \beta _{h1}{m^4 \over \lambda } \big(1-{\beta _1\over 2}s \big)^
{1+{2\gamma _{m1}\over \beta _1}}
    (-\2{1})\int _s^{\bar s}d\bar s\,
   \big(1-{\beta _1\over 2}\bar s \big)^{-2(1+{\gamma _{m1}\over
\beta _1})}  \cr
 &= {m^4 \over \lambda } {\beta _{h1}\over \beta _1+2\gamma _{m1}}
\left[ 1 - \Big(
{1-{\beta _1\over 2}s \over 1-{\beta _1\over 2}\bar s} \Big)^{1+{2
\gamma _{m1}\over \beta _1}}\right]  \cr
}\ee
$$
so that
$$
\bar h \bar m^4 =  h m^4 +
 {m^4 \over \lambda } {\beta _{h1}\over \beta _1+2\gamma _{m1}}
\left[ 1 - \Big(
{1-{\beta _1\over 2}s \over 1-{\beta _1\over 2}\bar s} \Big)^{1+
{2\gamma _{m1}\over \beta _1}}\right] \ .
\ee$$

Now that we have determined the RG running of all the relevant
parameters, we
can write down the improved \ep\ in the leading-log order
by inserting them
into the `boundary' function. The `boundary' function
to the leading-log
order is given by the tree potential $V^{(0)}$.  But we use here
the 1-loop
potential $V_1=V^{(0)}+V^{(1)}$ with $s$ set equal to zero,
since it gives in any case a better
approximation in the region in which $\ln(M_\phi ^2/\mu ^2)$
 is not so large (and
to keep it is harmless in the sense of leading log expansion).
Then the potential at $s=0$ is given simply by setting
$\mu ^2=M_\phi ^2$ {\it directly} in $V=V^{(0)}+V^{(1)}$ and then
replace
all the parameters there by the above obtained barred ones with
$\bar s=0$
substituted.  Thus the leading-log order \ep\ is found to be:
$$
V = {1\over 2}\bar m^2 \bar\phi ^2 +{1\over 4!}\bar\lambda \bar
\phi ^4 + \bar h\bar m^4 -{3\over 2}
 {1\over 64\pi ^2} \big(\1{2}\bar\lambda \bar\phi ^2+\bar m^2)^2
\eqn\eqresult
$$
with
$$
\eqalign{
\bar\phi  &= \phi  \ ,\cr
\bar\lambda  &=
   \lambda \left( 1-{3\lambda \over 32\pi ^2}\ln{\sm\over \mu ^2}
\right)^{-1} \ ,\cr
\bar m^2 &=
   m^2\left( 1-{3\lambda \over 32\pi ^2}\ln{\sm\over \mu ^2}
\right)^{-{1\over 3}} \ ,\cr
\bar h \bar m^4 &=  h m^4 +  \1{2}{m^4 \over \lambda }
\left[ 1 -
   \left( 1-{3\lambda \over 32\pi ^2}\ln{\sm\over \mu ^2} \right)^{
{1\over 3}}\right]\ .\cr
}\ee
$$
This agrees with the result by Kastening \refmark{\Kast} aside from
the next-to-leading log
terms and the $\phi $-independent constant
terms. [Note that the singularity at $\lambda =0$ is automatically
absent here contrary to Kastening.]
An important point in this calculation is to demonstrate explicitly
that we need not even to rewrite the effective potential in terms of
the chosen
variables $s, x, z$. We have used those variables just to see the
correctness of our \ep\ to a certain order in the leading-log series
expansion. In practice we can simply substitute the running barred
parameters directly in the potential without
rewriting it in terms of $s, x, z$.

We add a remark on the numerical applications.
In cases in which there are
many coupling constants and masses (masses should be of the same
order), it
is necessary to carry out the calculation using computer. In such
cases of
numerical work, the above process of changing the variable from
$t$ to $s$
in solving the RG running is quite extraneous. (Doing so introduces
unnecessary complications.) The solution in the form
\eqSOLTWO\  before doing that already gives the final answer:
We first solve the RG running of
$\bar \lambda (t), \, \bar\phi (t)/\phi , \, \bar m^2(t)$ \etc,
for a given set of
initial coupling constants and mass parameters at $\mu $
(in which we do not yet need to specify $\phi $).
At the same time as we vary
the parameter $t$ in solving these differential equations, we can find
the corresponding $\phi $ by $\bar s(t)=0$; \ie,
$$
\phi ^2={2\over \bar \lambda (t)}\Big({\bar\phi (t)\over \phi }\Big)^
{-2}
   \big( e^{2t}\!\mu ^2 - \bar m^2(t) \big) \ .
\ee$$
[Note that the RHS does not depend on $\phi $.]
Then putting this value of $\phi $ and the running parameters into the
RHS of \eqSOLTWO, we obtain the \ep\ at that point $\phi $.  Namely
the \ep\ is
obtained simultaneously as we solve the RG running of the parameters.
Moreover, since we make no further approximation in solving the RG
running
of the barred quantities in this process, the obtained \ep\
satisfies exactly the RGE with given $(L+1)$-loop coefficient
functions.

Our procedure described in this paper is applicable
to any complicated system
{\it provided that the relevant mass scales are essentially unique.}
To explain this, let us now consider the general situation.
Generically,
if a system consists of several particles labeled by $j\
(j=1,2,\cdots ,n)$,
the \ep\ contains logarithm factors of the form:
$$
\lambda _j\ln{M_j^2(\phi )\over \mu ^2}\ ,
\eqn\eqLOGFAC
$$
where $M_j(\phi )$ is a mass of the $j$-th particle on the
background in which
scalar fields take VEV's $\phi =(\phi _1,\cdots \phi _n)$, and takes
the form
$$
M_j^2(\phi ) = \sum_i\lambda _{ji}\phi _i^2  + m_j^2 \ .
\ee$$
Here $\lambda _j$ and $\lambda _{ji}$ are (certain linear
combinations of)
coupling constants and $m_j$ is the mass of $j$-th particle in the
absence of
scalar field background.  First problem we
encounter here is which log-factor among these we should choose as
the $s$ variable with which we sum up the leading log,
next-to-leading log,
$\cdots$ terms. The best choice would be to take a particle whose
coupling constant
$\lambda _j$ is the largest; namely, calling that particle by label
$j=0$,
we take the corresponding log-factor as the
$s$ variable: $s\equiv \lambda _0\ln(M_0^2(\phi )/\mu ^2)$.
Then all the other log-factors are rewritten in the form
$$
\lambda _j\ln{M_j^2(\phi )\over \mu ^2} = {\lambda _j\over
\lambda _0}s + u_j,
\eqn\eqU
$$
with introducing  new variables
$$
u_j \equiv  \lambda _j\ln{M_j^2(\phi )\over M_0^2(\phi )} \ .
\eqn\eqVARU
$$
Assume now that all the masses $M_j^2(\phi )$ here
in the presence of scalar
field background, are of the same order {\it independently} of
the background $\phi $.
This is the situation which we meant in the above by
``relevant mass scales are essentially unique".
[This happens, for instance, if we are considering the \ep\ as
a function of a single scalar field $\phi $ with other VEV's set
equal to zero,
and the `bare' masses $m_j$ (and the
coupling constants $\lambda _j$) are all of the
same order; indeed in such a case, $M_j^2(\phi )$'s take the form
$\lambda _j\phi ^2+m_j^2$ and are of the same order as
$\lambda _0\phi ^2+m_0^2$ independently of the value of $\phi $.]
Then, these variables $u_j$ in \eqVARU\ remain always of the order
$O(\lambda _j)\siml O(\lambda _0)$ at most, and therefore
all the log-factors \eqLOGFAC\ can be treated essentially as $s$,
(or $(\lambda _j/\lambda _0)s$ more precisely,)
since the differences $u_j$ in Eq.\eqU\ are higher order (in
the leading-log series expansion) than the first $s$ term.

The $L$-th power terms in $s$ and $u_j$'s come
from $L$-(or higher-)loop
contributions to the \ep. But the variable $s$ among them can be
set equal to
zero when we obtain the boundary function. The other variables $u_j$
remain as they are. However, since $u_j$ are of $O(\lambda _0)$
under the above constraints, the $(L+1)$-loop or higher-loop
contributions
to the \ep\ after setting $s=0$,
$\sum_{\ell=L+1}^\infty V^{(\ell)}\big\vert_{s=0}$, can be of
the order
$\lambda _0^L$ at most. Therefore the previous order-counting
argument of the leading-log series expansion
remains unchanged; namely, the $L$-th-to-leading-log exact
boundary function can be obtained by the $L$-loop potential $V_L$
simply by
setting $\mu ^2=M_0^2(\phi )$ (\ie, $s=0$). The runnings of the
barred quantities,
$\bar\lambda _j, \bar m_j^2$ \etc, substituted there can
of course be computed correctly
by using the $(L+1)$-loop RGE coefficient functions.

This argument also clarifies the problem which occurs when the above
constraints are not met.
In such a case the variables $u_j$ in \eqVARU\
no longer remain small depending on the
region of $\phi _j$'s. For instance, even when we
are discussing the \ep\ of a single scalar field, $\phi _j=\phi $,
if a particle is massless, $m_j=0$, then
the corresponding $u_j$ is given by
$\lambda _j\ln\big(\lambda _j\phi ^2/(\lambda _0\phi ^2+m_0^2)\big)$.
This is of $O(\lambda _j)\siml O(\lambda _0)$ for large $\phi $,
$\lambda _0\phi ^2\geq m_0^2$, but
becomes very large for small $\phi $
in the region $\lambda _0\phi ^2\ll m_0^2$.  So we have to keep any
higher
powers of such $u_j$
in our leading-log series expansion.
\foot{
This difficulty due to the presence of multi-scales has been noticed
by
several authors.\refmark{\Sher,\EJ}
}
 This implies that we cannot find a
good boundary function by the present procedure:
if the boundary function is calculated by $L$-loop potential $V_L$
with $s$ set equal to zero, then it is correct only up to
$L$-th power in the $u_j$ variables, and so it becomes completely
unreliable
in the small $\phi $ region  $\lambda _0\phi ^2\ll m_0^2$.

This problem, of course, stems from our careless treatment of
different mass scales by a single scale parameter $\mu $.
It turns out to be
overcome by a proper use of decoupling
theorem or the so-called effective field theory. Renormalization
group
equation in fact contains this notion of effective field theory
in a very natural form. Using this we can still have a simple
procedure
of improving \ep\
for completely general system. This will be given
in a separate paper.\refmark\BKMNa

\refout

\bye